# Ferrofluid Droplet Behavior on Gradient Surfaces Inside a Uniform Magnetic Field

Mojtaba Edalatpour[‡1], Khalid Eid [†], Andrew Sommers[††]

[‡]Department of Mechanical Engineering, Virginia Tech, Blacksburg, VA 24061, United States

[†]Department of Physics, Miami University, Oxford, OH 45056, United States

[††]Department of Mechanical and Manufacturing Engineering, Miami University, Oxford, OH 45056, United States

---

[1]Corresponding author: edalatm@vt.edu (M. Edalatpour)

Emails: edalatm@vt.edu; eidkf@miamioh.edu; sommerad@miamioh.edu

# ABSTRACT

The spontaneous motion of liquid droplets on solid surfaces is the result of an unbalanced surface tension force, which is sometimes called the "Marangoni effect". This can be triggered by either a difference in surface temperature or a heterogeneity in the topography or chemistry of the surface passively or actively. The imbibition of liquid within capillary tubes, horizontal ice wicking on either hydrophilic or hydrophobic substrates, and inkjet printing for example are just some classic illustrations of where the Lucas-Washburn equation can predict droplet behavior characteristics fairly well. In contrast, this study reveals an example of droplet behavior not previously studied that is not well-predicted by the Lucas-Washburn equation, namely the motion of ferrofluid droplets in the presence of uniform magnetic field. When a ferrofluid droplet is horizontally exposed to an external uniform magnetic field on a biphilic surface tension gradient in the shape of a wedge, it appears to violates the Lucas-Washburn equation which predicts that droplet travel distance should scale with the square root of time $(\text{i.e. } l \sim t^{1/2})$. Rather, our experimental results suggest that the movement of the ferrofluid droplet is slower following the relationship, $(l \sim t^{1/3})$. Furthermore, due to the relatively high viscosity of water-based ferrofluid droplets, we observed that at the beginning of the motion, the visco-capillary effect dominates the effects of the magnetism, and the droplets tend to follow the well-known relationship, $(l \sim t^{1/10})$. This initial stage of droplet spreading is known as "Tanner's Law".

# INTRODUCTION

Ferrofluids are colloidal suspensions of either γ-$Fe_2O_3$ (maghemite) or $Fe_3O_4$ (magnetite) or Co•$Fe_2O_4$ (cobalt ferrite) subsumed under the category of complex fluids. These binary solid-liquid systems contain single domain iron oxide magnetic particles whose sizes are generally in the range of 5-20 nm [1]. They have a variety of practical applications especially with regards to diagnostic tests and experimental therapies for targeting malignant diseases which often require the use of ferrofluids to be successful. For example, the human body contains an innumerable number of blood vessels which act like capillary tubes circulating blood all around the body. In magnetic drug delivery, magnetic nanoparticles are injected into the blood stream and external magnetic fields are then employed to steer these nanoparticles towards diseased locations [2-12]. In 1920, Professor Edward W. Washburn theoretically postulated that the length of travel of a simple liquid flow ($l$) inside a horizontal capillary under the effect of its own capillary pressure (i.e. $P_s = \frac{2\gamma}{r} \cos\theta$) might be proportional to the effective radius of the capillary ($r$), surface contact angle ($\theta$), surface tension ($\gamma$), viscosity ($\mu$), and the square root of time ($t$). Through experimentation using a handful of simple Newtonian liquids, namely water and mercury, he subsequently demonstrated that his theoretical equation was valid. The contemporary form of his well-known equation can be written as [13]:

$$l(t) = \sqrt{\frac{r\gamma \cos\theta}{2\mu}} \; t^{1/2} \qquad (1)$$

Nevertheless, this model had already been briefly postulated in 1918 by Richard Lucas. Thus, Eqn. (1) is often referred to as the Lucas-Washburn equation [14]. Three distinct stages are often associated with the spontaneous vertical capillary wetting of simple Newtonian liquids, namely (1) inertia regime taking place at the beginning of the wetting, (2) Lucas-Washburn regime where Eqn. (1) works well, and (3) the very end of the wetting where gravitational forces dominate over the capillary forces [15]. Although Eqn. (1) is simple, straightforward, and can explain a myriad of macroscopic wetting phenomena, it has some inherent shortcomings. First, the Lucas-Washburn equation appears to be unable to explain the behavior of complex solid-liquid binary systems like ferrofluids and nanofluids, in which nanoscale particle-to-particle interactions including Brownian motions exist. Second, the Lucas-Washburn equation neglects to take volumetric body forces like magnetism and gravity into consideration. Lastly, this equation is unable to predict what would happen on a heterogeneous capillary surface possessing a topographically and/or chemically-modified surface wettability where the static and dynamic contact angles are variable with respect to both time and planar location. For these reasons, departure from the Lucas-Washburn equation for simple Newtonian liquids on homogeneous surfaces has already been observed and theoretically modeled [15-16]. Nevertheless, a rigorous theoretical explanation, along with an experimental study, of the impact of using complex liquids and magnetic fields on surface tension gradients from the perspective of the Lucas-Washburn equation remains an outstanding issue in the field of droplet wetting. In this work, the droplet spreading behavior of commercially available, water-based ferrofluid (EMG 700, Ferrotec Corp.) on a surface tension gradient is compared to the well-known Lucas-Washburn equation both in the absence and presence of an external uniform magnetic field where the magnetic field lines are aligned either parallel or perpendicular to the gradient direction.

**Experimentation**

These plates were first cleaned in a solution of Alconox and water of arbitrary molarity immediately after removing the protective cover to remove any residue left on the surface due to the protective cover. The surface was then gently wiped with a clean Kimwipe during the Alconox soaking followed by rinsing with copious amounts of water to clean the surface from the Alconox powder thoroughly and to remove any organic and/or non-organic deposits from the surface. In the next step, both the plate and the accompanying wedge-angle shadowmask, which was made out of a conventional aluminum sheet, were clamped together and placed inside a thermal evaporation chamber in preparation for physical vapor deposition (PVD). Note that the wedge angles, i.e. $(\psi_{max} = 17°, \psi_{min} = 10°)$, were selected arbitrarily to provide a variety of angles in order to evaluate their potential impact on the spreading behavior of the ferrofluid droplets. The physical vapor deposition of copper nanoparticles (Cu) on the aluminum was performed at $5 \times 10^{-5}$ Pa. Once the thickness of the copper layer which was selected arbitrarily, reached approximately 120 nm on the aluminum plate, the process was completed. After the surface was taken out of the chamber, the surface was immersed in a 0.1 M heptadecafluoro-1-decanethiol (HDFT) solution for approximately 6-8 minutes allowing the surface to form a HDFT self-assembled monolayer (SAM) on the Cu regions. (Note: Previous research has shown that the HDFT self-assembly process favors the Cu region over the Al [17]). This rendered the Cu region considerably less hydrophilic than the Al region thus enabling EMG 700 ferrofluid droplets ($\rho = 1.29 \text{ g/cc}, \mu = 5\text{cP}, \phi = 5.8\%, M_s = 325 \text{ G}, \gamma = 51 \text{ mN/m}$) to spontaneously travel down the gradient along the wedge. A representative image of one of the produced samples is shown in the supporting information.

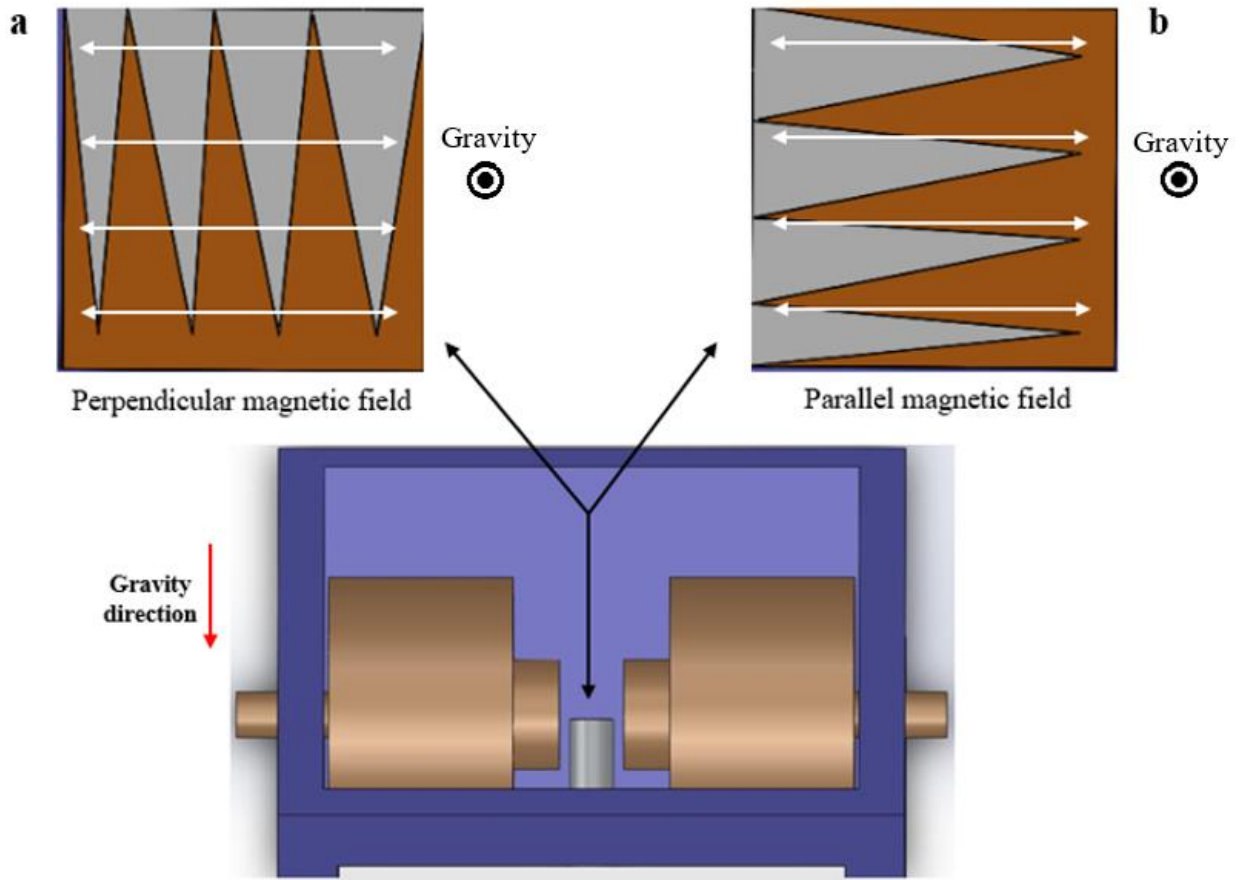

FIG. 1. Schematic of the experimental setup: (a) the surface horizontally sits between of the horizontal electromagnet poles. The electromagnetic field denoted by white double arrows, is perpendicular to both gravity and the direction of the surface tension gradient, (b) the surface is horizontally placed between the electromagnet poles. The electromagnetic field delineated by white double arrows, is perpendicular to gravity, while it is parallel to the direction of the surface tension gradient.

Fig. 1 illustrates the experimental setup that was used to study the spontaneous motion of ferrofluid droplets placed on the surface tension gradient. In Fig. 1(a), the surface was horizontally placed between the two large cylindrical electromagnet poles (Alpha Scientific Systron) such that the surface tension gradient direction was perpendicular to the horizontal electromagnetic field (white double line arrows), while gravity was perpendicular to both the surface tension gradient and magnetic field lines. In Fig. 1(b), the surface was again horizontally located between the two cylindrical electromagnet poles such that in this case the surface tension gradient direction was parallel to the horizontal electromagnetic fields (white double line arrows), while gravity was perpendicular to both the surface tension gradient and magnetic field lines.

# RESULTS AND DISCUSSION

Deriving the spanwise Stokes equation in Cartesian coordinates as well as the continuity equation for a one-dimensional, incompressible fluid in the presence of volumetric magnetic forces on a surface tension gradient results in:

$$\frac{\partial u}{\partial x} = 0 \tag{2}$$

$$\frac{\partial p}{\partial x} = \mu \left(\frac{\partial^2 u}{\partial y^2}\right) + F_{\text{magnet}} \tag{3}$$

where $u$ is the velocity in span-wise direction, $p$ is the capillary pressure, $\mu$ is the magnetoviscosity of the ferrofluid droplets, and finally $F_{\text{magnet}}$ is the magnetic field applied to the ferrofluid droplets which can be stated namely as [18]:

$$F_{\text{magnet}} = \left(\frac{2\pi r_{\text{np}}^3}{3}\right) \frac{\mu_0 \chi}{(1 - \chi/3)} H \cdot \nabla H \tag{4}$$

In Eqn. (4), $r_{\text{np}}$ is the radius of magnetic particles, $\mu_0$ is the magnetic permeability, $\chi$ is the magnetic susceptibility, and $H$ is the external uniform magnetic field strength. If the magnetic field reaches the saturation magnetization ($M_s$) of the ferrofluid droplet, $H \cdot \nabla H$ needs to be replaced by $M_s \cdot \nabla H$. For short times where $t \ll 1\ s$, the injected ferrofluid droplet tends to first reach its steady state by spreading on the surface. Assuming that the droplet's volume can be modeled as a spherical cap ($V \sim l^2 h$) while the droplet curvature is $k \sim h/l^2$, Eqn. (3) can be rewritten using visco-capillary scaling principles such that:

$$\gamma \left(\frac{V^3}{l^9}\right) \sim \frac{\mu l}{t} + \mu_0 h^2 CH \tag{5}$$

in which $C$ is the current density. Rearranging Eqn. (5) yields:

$$l(t) \sim \left(\frac{t}{\mu}(\gamma V^3 - \mu_0 h^2 CH l^9)\right)^{1/10} \tag{6}$$

However, because $\mu_0 h^2 CH l^9 \sim 0$, Eqn. (6) can be rewritten in a more simplified form such that:

$$l(t) \sim \left(\frac{\gamma V^3 t}{\mu}\right)^{1/10} \tag{7}$$

Eqn. (7) is the well-known equation commonly referred to as "Tanner's law".

Once the droplet reaches its final steady-state shape on the surface, writing the scaling law for Eqn. (3) while considering the volumetric magnetic force, one obtains:

$$\frac{\gamma}{rl} \sim \mu \left(\frac{\dot{l}}{r^2}\right) + \mu_0 CH \qquad (8)$$

Eqn. (8) demonstrates that the driving capillary pressure gradient for the droplet at steady-state on the surface hinges upon both viscous friction based upon the velocity of the ferrofluid droplet, and the magnetic force. Eqn. (8) can be simplified as:

$$l\dot{l} \sim \frac{\gamma r}{\mu}(1 - \frac{\mu_0 CHrl}{\gamma}) \qquad (9)$$

Taking the first derivative of Eqn. (9) with respect to the radius ($r$) and setting $r = r_L$ renders:

$$r_l \sim \frac{\gamma}{2l\mu_0 CH} \qquad (10)$$

The length of the capillary using Eqns. (7-10) at longer times can thus be found as:

$$l(t) \sim \left(\frac{\gamma^2 t}{\mu \mu_0 CH}\right)^{1/3} \qquad (11)$$

This scaling argument is shown to be in excellent agreement with the experimental results presented in Figs. 2 and 3 where the length of capillary wetting of ferrofluid droplets at short times follows $t^{1/10}$, while at longer times it obeys $t^{1/3}$ for various magnetic field intensities and two distinct orientations. Germane movies are available in the supporting information). Note that although some of the experimental results posed in Fig. 2 were taken outside of the electromagnetic field, due to the magnetic particles in the ferrofluid droplets possessing charges and random dipole moments, the length of the capillary wetting still perfectly follows $t^{1/3}$. The capillary wetting length according to Eqn. (11) is also dependent upon $\psi$. In other words, the distance that droplets can travel on the surface is greater on wedges possessing smaller wedge angles. Lastly, Eqn. (11) also shows that the capillary length is directly proportional to the surface tension of ferrofluid droplets while it has an inverse relation with the magnetoviscosity of these droplets.

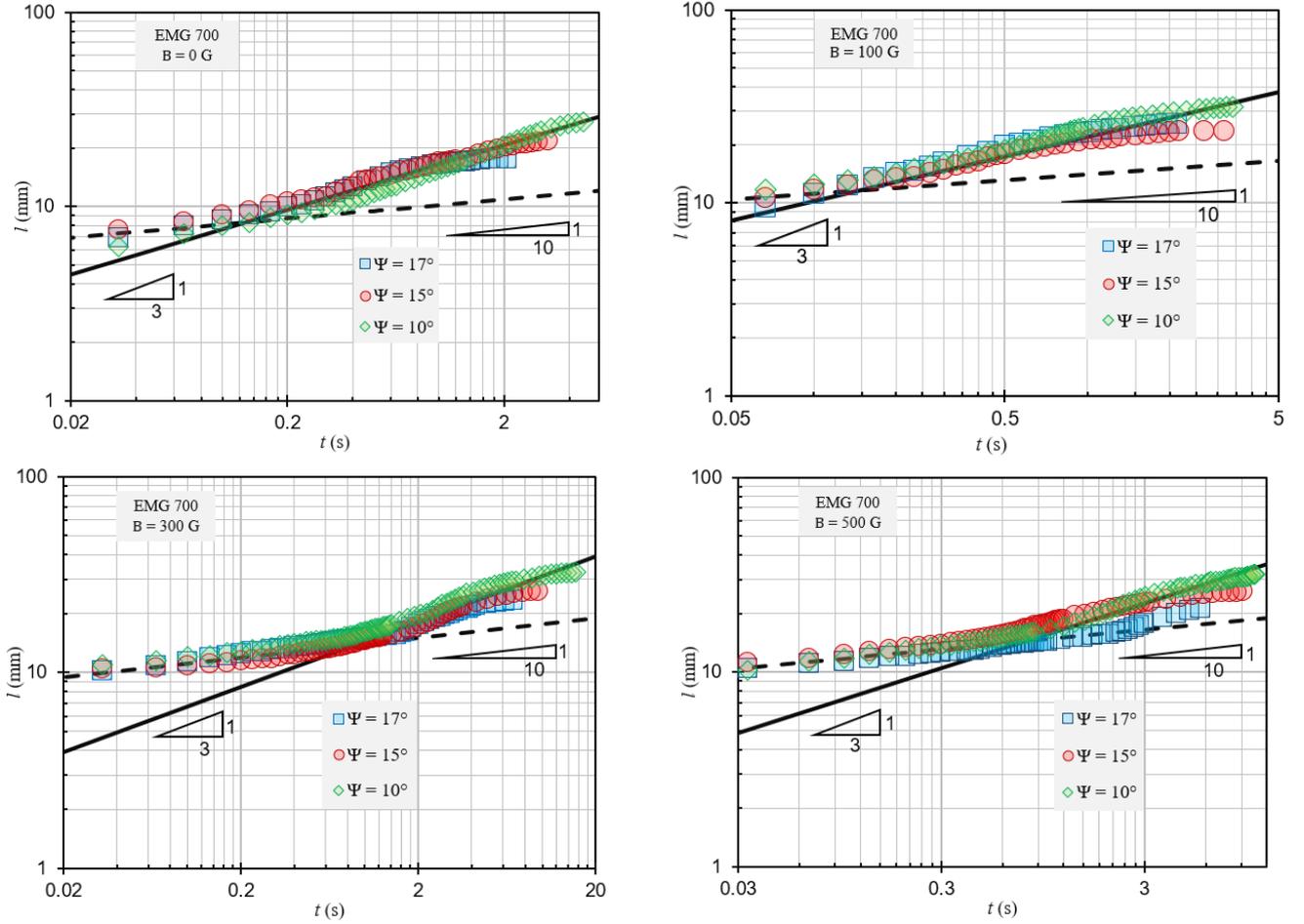

FIG. 2: Spontaneous motions of EMG 700 ferrofluid droplets in the parallel orientation in both the absence and presence of magnetic fields of varying strength.

These results also show that the magnetoviscosity of the ferrofluid droplet in the perpendicular orientation is greater than that of the parallel orientation (see Fig. 3), confirming what has been reported elsewhere [20]. Droplets travel much smaller overall distances when in the perpendicular orientation as compared to the parallel orientation. Fig. 3 also illustrates that not only does the distance travelled by the ferrofluid droplet on the surface tension gradient at short and long times follow the scaling argument derived in Eqns. (7) and (11), but it also observes the inverse relation shown in Eqn. (11) with respect to the magnetoviscosity of the ferrofluid droplets. Fig. (3) reveals that the droplet motion on $10°$ wedge angle is thoroughly blocked by the magnetic field for the whole range of magnetic fields investigated in this research.

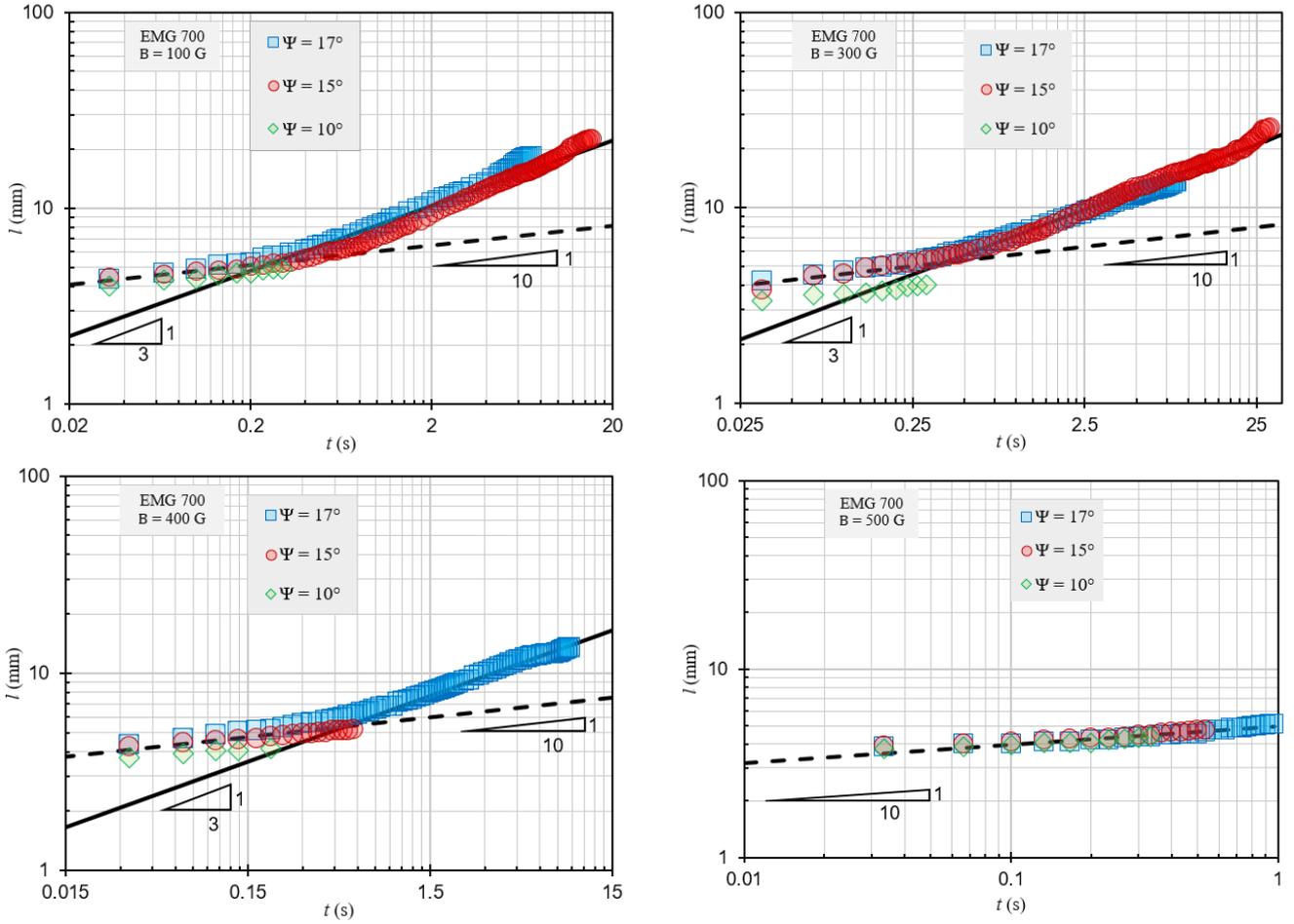

FIG. 3: Spontaneous motions of EMG 700 ferrofluid droplets in the perpendicular orientation in the presence of magnetic fields of varying strength.

In addition, Fig. (3) shows that the droplet is "pinned" to the surface at 500 G following the initial deposition and spreading, and as such does not spontaneously travel down the length of the wedge as in the previous cases. Interestingly, the deposited ferrofluid droplet can still reach its steady-state configuration by increasing its surface area exposed to the surroundings (thereby minimizing its total free energy); however, after this has occurred (on the order of $10^{-1}$ s), the droplet appears to be "pinned" to the surface showing no tendency to travel along the wedge in the direction of the gradient. This reproducible control over the shapes, deformations, and movements of these ferrofluid droplets on biphilic surfaces offers a vast promising applications in optics, smart sensors and actuators, and microfluidics pumps. Eventually, to further strengthen the validity of the rigorous mathematical derivations along with the experimental results, experiments were also repeated using oil-based ferrofluid droplets (APG J12, Ferrotec Corp.) ($\rho = 1.11 \text{ g/cc}, \mu = 40\text{cP}, \phi = 1.4\%, M_s = 220 \text{ G}, \gamma = 32 \text{ mN/m}$), where its results can be found in the supporting information.

## CONCLUSIONS

The impact of an external uniform magnetic field on the spreading behavior of a non-Newtonian, water-based ferrofluid droplets on a horizontal surface tension gradient in two distinct orientations (parallel and perpendicular) with respect to gradient directions was examined. It was observed that the motion of the ferrofluid droplets followed the $l \sim t^{1/10}$ (visco-capillary) relationship at short times, while it tended to follow a $l \sim t^{1/3}$ relationship at longer times in apparent contradiction to the Lucas-Washburn equation for simple Newtonian liquids in the absence of volumetric magnetic body forces. Since the magnetoviscosity of the ferrofluid droplets was shown to be greater in the perpendicular orientation, ferrofluid droplets in this orientation generally did not travel as far along the wedge (i.e. $l_\perp < l_\parallel$). Moreover, in certain cases (i.e. 500 G), the magnetic field was observed to actually "pin" the droplet to the surface in the perpendicular orientation preventing its motion down the gradient altogether. Finally, although a handful of cases were observed where the motion of the droplet on the surface (following deposition) was blocked by the magnetic field, the magnet was unable to prevent the initial wetting behavior of the droplet on the surface as the droplet minimized its total free energy (Gibbs free energy) en route to a thermodynamically favored configuration.

## ACKNOWLEDGEMENT

The first author (M.E.) is grateful of Abbas Khadivar, Mark Fisher, Dr. S.F. Ahmadi, and Zeeshan Ali.